\def\ra{\rangle}
\def\la{\langle}
\def\be{\begin{equation}}
\def\ee{\end{equation}}
\def\ba{\begin{array}}
\def\ea{\end{array}}

\documentclass[prl,showpacs,twocolumn,amsmath]{revtex4}
\usepackage{mathrsfs}
\usepackage{amssymb}
\usepackage{pifont}
\usepackage{epsfig,subfigure,dsfont,amsthm,amsbsy,mathrsfs,amscd}
\def\qed{\leavevmode\unskip\penalty9999 \hbox{}\nobreak\hfill
     \quad\hbox{\leavevmode  \hbox to.77778em{%
               \hfil\vrule   \vbox to.675em%
               {\hrule width.6em\vfil\hrule}\vrule\hfil}}
     \par\vskip3pt}

\begin{document}
\title{ Note on Product-Form Monogamy Relations for Nonlocality and Other Correlation Measures}
\author{Tinggui Zhang$^{1,*}$, Xiaofen Huang$^{1}$, Shao-Ming Fei$^{2,3}$}

\affiliation{ $^1$School of Mathematics and Statistics, Hainan
Normal University, Haikou, 571158, China\\
 \footnotesize \small
$^{2}$ Max-Planck-Institute for Mathematics in
the Sciences, Leipzig 04103, Germany \\
 \footnotesize\small $^{3}$ School of Mathematical Sciences, Capital Normal University, Beijing 100048,
 China\\
  \footnotesize $^*$ E-mail: tinggui333@163.com}

\begin{abstract}
{\bf Abstract}: The monogamy relations satisfied by quantum
correlation measures play important roles in quantum information
processing. Generally they are given in the summation form. In this
note, we study monogamy relations in product form. We present
product-form monogamy relations for Bell nonlocality for three-qubit
and multi-qubit quantum systems. We then extend our studies to other
quantum correlations such as concurrence.
\end{abstract}


\maketitle

\section{Introduction}
Quantum technologies radically change the landscape of modern
communication and computation. Quantum correlations, such as quantum
discord \cite{howh,kmab}, quantum entanglement \cite{nils,rpmk},
quantum steering \cite{hmsj,cdsp}, quantum nonlocality
\cite{hbsp,bncd} have shown to be useful resources in many quantum
information processing tasks like quantum cryptography \cite{gnrg},
quantum metrology \cite{fadp}, quantum illumination \cite{cwsp}.
Various measures have been proposed to quantify these correlations
\cite{gatr}. Generally these quantifiers are very difficult to
calculate, except for some specific cases like two-qubit states
\cite{rhph,vcjk,luos,acsc,tzhy}.

A successful and secure quantum network relies on quantum
correlations distributed and shared over many sites \cite{hjki}.
Different kinds of multipartite quantum correlations have been
considered as valuable resources for various applications in quantum
communication tasks. A key property is that such quantum
correlations cannot be freely shared among the multipartite systems.
The monogamy relation states that the more two systems correlated,
the less with the rest systems. It was first shown in \cite{vcjk} that the
bipartite entanglement measure concurrence $C$ of the reduced states
$\rho_{AB}$ and $\rho_{AC}$ of a three-qubit state $\rho_{ABC}$
satisfies the Coffman-Kundu-Wootters (CKW) relation,
$C(\rho_{A|BC})\geq C(\rho_{AB})+C(\rho_{AC})$, where
$C(\rho_{A|BC})$ stands for the concurrence between subsystems $A$
and the remaining subsystems $BC$. The CKW relation means that the
sum of the individual pairwise entanglement between $A$ and $B$ and
$C$ cannot exceed the entanglement between $A$ and the remaining
parties together. Since then there have been many papers focused on
such monogamy or polygamy relations for quantum entanglement
\cite{tjfv,xzsf,liuf,jzx,gyu}. In \cite{clsd} the generalized
summation-form monogamy relations for any valid entanglement measure
have been investigated. Monogamy relations have also been studied
for quantum discord \cite{asga,yknz}, quantum steering
\cite{mdam,czsc}, Bell nonlocality \cite{vsng,btfv,pktp,scmj},
indistinguishability \cite{mkdk}, coherence \cite{crmp} and other
nonclassical correlations \cite{scll}.

For tripartite quantum systems, monogamy relations have the following general (trade-off) form,
\begin{eqnarray}\label{eq1}
Q^\alpha(\rho_{AB})+Q^\alpha(\rho_{AC}) \leq Q^\alpha(\rho_{A|BC})
\end{eqnarray}
or
\begin{eqnarray}\label{eq2}
Q(\rho_{AB})+Q(\rho_{AC}) \leq K
\end{eqnarray}
for some bipartite quantum correlation measure $Q$ and positive real
numbers $\alpha$ and $K$, where $Q(\rho_{A|BC})$ stands for the
correlation between subsystems $A$ and the remaining subsystems
$BC$, the vertical bar is the familiar notation for the bipartite
split, $Q(\rho_{AB})$ ($Q(\rho_{AC})$) represents the bipartite
correlation the reduced state $\rho_{AB}$ ($\rho_{AC}$) of the
tripartite state $\rho_{ABC}$.

Generally (\ref{eq1}) does not hold for $\alpha=1$ for many
correlation quantifiers like the geometric measure of discord
\cite{asga}). However, for any given $Q$, (\ref{eq1}) holds for
sufficient large $\alpha$ \cite{jzx,gyu}). For example, it has been
shown that the $\alpha$th ($\alpha \geq 2$) power of discord for
3-qubit pure states \cite{asga,yknz}, the $\alpha$th ($\alpha \geq
2$) power of concurrence and the $\alpha$th ($\alpha \geq \sqrt{2}$)
power of entanglement of formation for N-qubit states do satisfy the
monogamy relations \cite{xzsf}.

And for Eq.(\ref{eq2}), there are also some examples for
entanglement \cite{liuf} and  Bell nonlocality \cite{scmj} of three
qubit states. The well-known CHSH-Bell \cite{jcmh} operator is given
by, \be\label{chsh22} \mathcal{B}=A_1\otimes B_1+A_1\otimes
B_2+A_2\otimes B_1-A_2\otimes B_2, \ee where
$A_i=\vec{a}_i\cdot\vec{\sigma}$, $B_j=\vec{b}_j\cdot \vec{\sigma}$,
$\vec{a}_i$ and $\vec{b}_j$ are three-dimensional real unit vectors,
$\vec{\sigma}=(\sigma_1,\sigma_2,\sigma_3)$ with $\sigma_1$,
$\sigma_2$ and $\sigma_3$ the standard Pauli matrices, $i,j=1,2$.
The CHSH inequality says that $|\la \mathscr{B}(\rho)\ra |\leq2$,
where $\la \mathscr{B}(\rho) \ra=tr(\rho\mathcal{B})$ is the mean
value of the Bell operator $\mathcal{B}$ associated with the state
$\rho$. For tripartite systems, the Bell violations among the
reduced bipartite subsystems have been investigated \cite{hqsf},
$$
\langle \mathscr{B}(\rho_{AB})\rangle^{2}+ \langle \mathscr{B}(\rho_{AC}) \rangle^{2}+ \langle \mathscr{B}(\rho_{BC})\rangle^{2}\leq 12.
$$
It has been further shown that
\begin{eqnarray}\label{eq4}
\langle\mathscr{B}(\rho_{AB})\rangle^2+\langle\mathscr{B}(\rho_{AC})\rangle^2\leq 8.
\end{eqnarray}

Cheng et al \cite {scll} also considered monogamy relation in
summation form like
$\langle\mathscr{B}(\rho_{AB})\rangle^{\alpha}+\langle\mathscr{B}(\rho_{AC})\rangle^{\alpha}\leq
\langle\mathscr{B}(\rho_{A|BC})\rangle^{\alpha}$. However, it is not
true in general. For example, given the state $|000\rangle$, one has
$\langle\mathscr{B}(\rho_{AB})\rangle=\langle\mathscr{B}(\rho_{AC})\rangle
=\langle\mathscr{B}(\rho_{A|BC})\rangle=2$. The investigation on
such relations for three-qubit pure states has applications in the
study of bi-locality \cite {ngqm}, and can be easily extended to
other kinds of quantum correlations such as concurrence.

As for quantum uncertainty relations, there are both summation forms
\cite{whei,jahz,hpro} and product forms \cite{lmak}, which have
their own advantages. Since the monogamy relations of Bell
nonlocality in summation form equation (1) or the CKW type do not
exist, it is worthy of studying the possible product forms.

In this paper, we mainly investigate the product-form monogamy
relations for nonlocal correlation measures. They can not be derived
from the summation-form monogamy relations which even do not exist
in some cases. Concerning the relations among the summation and
product-form monogamy relations, we show in Fig. 1 the parameter
regions where different monogamy relations are associated. These
product-form monogamy relations we obtained can not be obtained from
summation-form ones by arithmetic geometric mean inequalities. They
are tighter than the ones derived from the summation-form relations
(if they exist).

\begin{figure}[ptb]
\includegraphics[width=0.4\textwidth]{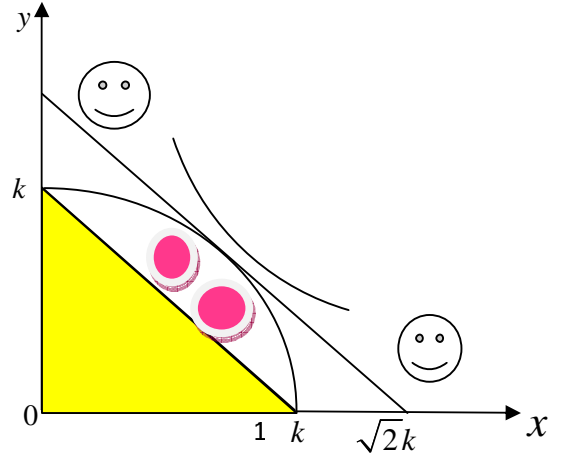}\caption{Yellow (pink) zone
stands for quantum correlations satisfying monogamy relations of the
form $x+y \leq k$ ($x^2+y^2 \leq k^2$). For Bell nonlocality, the monogamy
relations satisfy the product form, $xy \leq ck$ with
$\frac{1}{2}\leq c\leq1$, the area where the smiling face located.}
\label{Fig 1}%
\end{figure}
\setlength{\abovecaptionskip}{0.cm}

\section{product-form monogamy relations for Bell nonlocality of three qubit states}

A two-qubit state $\rho$ can be expressed as
$$
\rho=\frac{1}{4}(\mathbb{I}\otimes\mathbb{I}+\vec{a}\cdot\vec{\sigma}\otimes\mathbb{I}
+\mathbb{I}\otimes\vec{b}\cdot\vec{\sigma}+\sum_{n,m=1}^{3}t_{m,n}\sigma_{m}\otimes\sigma_{n}),
$$
where $\mathbb{I}$ is the identity operator. Let $T_{\rho}$ denote the
real matrix with entries given by $t_{nm}=Tr(\rho\sigma_{n}\otimes\sigma_{m})$.
Set $U_{\rho}\equiv T_{\rho}^{T}T_{\rho}$. Denote $\mu$ and $\tilde{\mu}$
the two greater eigenvalues of $U_{\rho}$ and $M_{\rho}\equiv\mu+\tilde{\mu}$.
Then $\langle\mathscr{B}(\rho_{AB})\rangle=2\sqrt{M_{\rho_{AB}}}$ \cite{rhph}.
Then, (\ref{eq4}) has the following simple form,
\begin{eqnarray}
\label{eq5}
M_{\rho_{AB}}+M_{\rho_{AC}}\leq 2.
\end{eqnarray}
Using the arithmetic geometric mean inequality, we have
\begin{eqnarray}\label{eq8}\sqrt{M_{\rho_{AB}}M_{\rho_{AC}}}\leq 1.\end{eqnarray}

Now consider general three-qubit pure states \cite{aale},
$$
|\psi_{ABC}\rangle=l_0|000\rangle+l_1e^{i\phi}|100\rangle+l_2|101\rangle+l_3|110\rangle+l_4|111\rangle,
$$
with $l_j \geq 0$ and $\sum_{j=0}^4l_j^2=1$.
Under the bipartition $A$ and $BC$, one has \cite{scll}, $M_{\rho_{A|BC}}=2-|\vec{a}|^2$, where
$\vec{a}=(2l_0l_1 cos\phi, 2l_0l_1 sin\phi, 2l_0^2-1)^T$ \cite{scmj}. Hence,
$M_{\rho_{A|BC}}=2-[4l_0^2l_1^2+4l_0^4+1-4l_0^2]=1-4l_0^2(l_1^2+l_0^2-1)\geq 1$.
In fact, under the bipartition $A$ and $BC$, $|\psi_{ABC}\rangle$ can be written as a
Schmidt bi-orthogonal sum,
$$|\psi_{A|BC}\rangle=\sum_{i=1}^2c_i|\phi_i\rangle|\theta_i\rangle.$$
By \cite{ngap}, one has
$\langle\mathscr{B}(|\psi_{A|BC})\rangle=2\sqrt{[1+[2(c_1c_2)]^2]}$, namely,
$M_{\rho_{A|BC}}\geq 1+[2(c_1c_2)]^2 \geq 1$.

Therefore, similar to the summation-form monogamy relation
(\ref{eq1}), for three-qubit pure stat $|\psi_{ABC}\rangle$ we have
the product-form monogamy relation by using (\ref{eq8}),
\begin{eqnarray}\label{eq7}
\sqrt{M_{\rho_{AB}}M_{\rho_{AC}}}\leq M_{\rho_{A|BC}}.
\end{eqnarray}

In fact, (\ref{eq7}) can be further tightened without the use of
(\ref{eq8}). We have the following result:

{\sf [Theorem]} For three-qubit pure state $|\psi_{ABC}\rangle$,
\begin{eqnarray}\label{thm}
M_{\rho_{AB}}M_{\rho_{AC}} \leq
\frac{1}{1+[C(\rho_{A|BC})]^2}M_{\rho_{A|BC}},
\end{eqnarray} where $C(\rho_{A|BC})$ is the concurrence of
$|\psi_{A|BC}\rangle$.

{\bf{Proof:}} It is direct to derive that \cite{scmj},
$M_{\rho_{AB}}=1+s_3^{AB}-s_3^{AC}-s_3^{BC}$,
$M_{\rho_{AC}}=1+s_3^{AC}-s_3^{AB}-s_3^{BC}$, where $s_3^{X}$
denotes the smallest eigenvalue of $U_{\rho_{X}}$. Thus,
$$
M_{\rho_{AB}}M_{\rho_{AC}}=[1-s_3^{BC}]^2-[s_3^{AB}-s_3^{AC}]^2.
$$
Moreover, since $s_3^{AB}-s_3^{AC}=|\vec{c}|^2-|\vec{b}|^2$, where
$\vec{b}=(2l_1l_3 cos\phi+2l_2l_4,-2l_1l_3sin\phi,1-2l_3^2-2l_4^2)$,
$\vec{c}=(2l_1l_2 cos\phi+2l_2l_4,-2l_1l_2sin\phi,1-2l_2^2-2l_4^2)$
\cite{scmj}, we obtain
\begin{eqnarray}\label{eqz1}
M_{\rho_{AB}}M_{\rho_{AC}}=[1-s_3^{BC}]^2-[|\vec{c}|^2-|\vec{b}|^2]^2.
\end{eqnarray}
Combining (\ref{eqz1}) and that $M_{\rho_{A|BC}}\geq 1+[2(c_1c_2)]^2
$, we get
\begin{eqnarray}\label{eqz2}
M_{\rho_{AB}}M_{\rho_{AC}} \leq
\frac{[1-s_3^{BC}]^2-[|\vec{c}|^2-|\vec{b}|^2]^2}{1+[2(c_1c_2)]^2}M_{\rho_{A|BC}}.
\end{eqnarray}
By the definition of concurrence for a bipartite pure state
$|\psi\rangle_{AB}$ \cite{auhl,runta,sasm},
$C(|\psi\rangle_{AB})=\sqrt{2(1-Tr(\rho^2_A))}$ with
$\rho_A=Tr_B(\rho_{AB})$ the reduced density matrix of
$\rho_{AB}=|\psi\rangle_{AB}\langle\psi|$ by tracing over the
subsystem $B$. We can obtain $C(|\psi_{A|BC}\rangle)=2(c_1c_2)$.
Theorem get a complete proof. \hfill \rule{1ex}{1ex}

Interestingly, we see that the tightened product-form monogamy
relation (\ref{thm}) for nonlocality depends also on the
entanglement concurrence. In \cite{ngqm}, the authors studied
bi-locality (It can be understood as the simplest three-point
quantum networks with one node in the middle) and showed that all
possible pairs of entangled pure stats can violate the so-called
``bi-locality" inequality. For arbitrary pairs of mixed two-qubit
states, the bi-locality inequality has the following form,
$$
S_{biloc}^{max}=2\sqrt{\xi_1\zeta_1+\xi_2\zeta_2}\leq 2,
$$
where, $\xi_1,~ \xi_2$ ($\zeta_1,~ \zeta_2$) are the first and second
large singular values of $U_{\rho_{AB}}$ ($U_{\rho_{AC}}$).
Due to $\langle\mathscr{B}(\rho_{AB})\rangle=2\sqrt{M_{\rho_{AB}}}$ \cite{rhph},
one has
$$
S_{biloc}^{max}\leq 2\sqrt{M_{\rho_{AB}}M_{\rho_{AC}}}.
$$
Hence, the violation of the bi-locality inequality implies that
$M_{\rho_{AB}}M_{\rho_{AC}} > 1$. Therefore, from (\ref{eq8}) all
the pairs of reduced density matrices from three-qubit states cannot
violate the bi-locality inequality.

Similar to the residual entanglement, we can also define
$$M_{\rho_{\overline{A}BC}}=M_{\rho_{A|BC}}-{M_{\rho_{AB}}M_{\rho_{AC}}}$$
to be the ``residual" quantum nonlocality of three-qubit pure states. For
the state $\alpha|000\rangle+\beta|111\rangle$, we have
$M_{\rho_{AB}}=M_{\rho_{AC}}=1$,
$M_{\rho_{A|BC}}=2-(|\alpha|^2-|\beta|^2)^2$, and thus
$M_{\rho_{\overline{A}BC}}=1-(|\alpha|^2-|\beta|^2)^2$. Therefore,
when $\rho_{ABC}$ is the GHZ state ($\alpha=\beta$),
$M_{\rho_{\overline{A}BC}}=1$. When $\rho_{ABC}$ is a fully separable state ($\alpha=0$ or $\beta=0$),
$M_{\rho_{\overline{A}BC}}=0$. Nevertheless, different from the
tangle of entanglement, $\tau_{ABC}=C^2_{\rho_{A|BC}}-C^2_{\rho_{AB}}-C^2_{\rho_{AC}}$ \cite{vcjk},
$M_{\rho_{\overline{A}BC}}$ is not invariant under the permutation of the qubits.

The monogamy relation (\ref{eq8}) can be generalized to multi-qubit
systems. It has been shown that \cite{scll},
\begin{eqnarray}
\label{eq6}M_{\rho_{AB}}+M_{\rho_{AC}}+M_{\rho_{AD}}+\cdots \leq n-1,
\end{eqnarray}
for any $n-$qubit pure or mixed state $\rho_{ABCD\cdots}$.
Using the generalized arithmetical geometric mean inequality
$\sqrt[n]{a_1a_2\cdots a_n} \leq \frac{a_1+a_2+ \cdots +a_n}{n}$,
we have for any $n-$qubit state $\rho_{ABCD\cdots}$
\begin{eqnarray}\label{eq9}
&\sqrt[n-1]{M_{\rho_{AB}}M_{\rho_{AC}}M_{\rho_{AD}}\cdots}\nonumber\\[2mm]
&\leq\frac{M_{\rho_{AB}}+M_{\rho_{AC}}+M_{\rho_{AD}}+\cdots}{n-1}\leq
1.\end{eqnarray}

In particular, an $n-$qubit pure states $|\psi_{ABCD\cdots}\rangle$ can be
viewed as a bipartite state under partition $A$ and $BCD\cdots$.
Therefore, according to the previous discussions on CHSH Bell
inequality violation of bipartite high-dimensional pure states,
we can obtain
\begin{eqnarray}\label{eq10}
\sqrt[n-1]{M_{\rho_{AB}}M_{\rho_{AC}}M_{\rho_{AD}}\cdots}
\leq M_{\rho_{A|BCD\cdots}}.
\end{eqnarray}
For multipartite quantum nonlocality, it is possible that the
product-form monogamy, like three-qubit, can be applied to more
complex quantum networks \cite{atmo,mcrr}.

\section{monogamy relations for other quantum correlations}

Our investigation on product-form monogamy relations for
non-locality can be extended to other quantum correlations.

Let us consider the entanglement measure concurrence
\cite{auhl,runta,sasm}. For three-qubit pure states
$|\psi\rangle_{ABC}$, the concurrence $\mathcal {C}_{AB}$ satisfies
\cite{vcjk}
$$
\mathcal{C}_{AB}^2=Tr(\rho_{AB}\widetilde{\rho}_{AB})-2\lambda_1\lambda_2,
$$
where $\lambda_1$ and $\lambda_2$ are the square roots of the
greater eigenvalues of matrix $\rho_{AB}\widetilde{\rho}_{AB}$,
$\rho_{AB}$ is the reduced density matrix of the qubit pair AB,
$\widetilde{\rho}_{AB}=(\sigma_2\otimes\sigma_2)\rho_{AB}^{\ast}(\sigma_2\otimes\sigma_2),$
with the asterisk denoting complex conjugation and $\sigma_2$ is the
Pauli matrix $ \left(
 \begin{array}{cc}
  0&-i\\
  i&0
 \end{array}
 \right).$

Similarly, the concurrence $\mathcal {C}_{AC}$ is given by
$$
\mathcal
{C}_{AC}^2=Tr(\rho_{AC}\widetilde{\rho}_{AC})-2\mu_1\mu_2.
$$
Therefore, \begin{eqnarray}\label{zh1}&&\mathcal {C}_{AB}^2\mathcal
{C}_{AC}^2\nonumber\\
&=&[Tr(\rho_{AB}\widetilde{\rho}_{AB})-2\lambda_1\lambda_2][Tr(\rho_{AC}\widetilde{\rho}_{AC})-2\mu_1\mu_2]\nonumber\\
&=&Tr(\rho_{AB}\widetilde{\rho}_{AB})Tr(\rho_{AC}\widetilde{\rho}_{AC})-2[Tr(\rho_{AB}\widetilde{\rho}_{AB})\mu_1\mu_2\nonumber\\
&+&Tr(\rho_{AC}\widetilde{\rho}_{AC})\lambda_1\lambda_2]+4\lambda_1\lambda_2\mu_1\mu_2.\end{eqnarray}

On other hand, the concurrence $\mathcal{C}_{A(BC)}$ between partition $A$ and $BC$ has the form, $\mathcal{C}_{A(BC)}^2=Tr(\rho_{AB}\widetilde{\rho}_{AB})+Tr(\rho_{AC}\widetilde{\rho}_{AC})$ \cite{vcjk}).
Therefore, we have
\begin{eqnarray}\label{zh2}&&
\mathcal{C}_{A(BC)}^2\nonumber\\&=&Tr(\rho_{AB}\widetilde{\rho}_{AB})+Tr(\rho_{AC}\widetilde{\rho}_{AC})\nonumber\\
&\geq&2\sqrt{Tr(\rho_{AB}\widetilde{\rho}_{AB})Tr(\rho_{AC}\widetilde{\rho}_{AC})}\nonumber\\
&=& 2\{\mathcal {C}_{AB}^2\mathcal
{C}_{AC}^2+2[Tr(\rho_{AB}\widetilde{\rho}_{AB})\mu_1\mu_2\nonumber\\&+&Tr(\rho_{AC}\widetilde{\rho}_{AC})\lambda_1\lambda_2]-4\lambda_1\lambda_2\mu_1\mu_2\}^{\frac{1}{2}}\nonumber\\
&=& 2\{\mathcal {C}_{AB}^2\mathcal
{C}_{AC}^2+2[(\lambda_1^2+\lambda_2^2)\mu_1\mu_2\nonumber\\&+&(\mu_1^2+\mu_2^2)\lambda_1\lambda_2]-4\lambda_1\lambda_2\mu_1\mu_2\}^{\frac{1}{2}}\nonumber\\
&\geq&2[\mathcal {C}_{AB}^2\mathcal
{C}_{AC}^2+4\lambda_1\lambda_2\mu_1\mu_2]^{\frac{1}{2}}\nonumber\\
&=&2[\mathcal
{C}_{AB}^2\mathcal{C}_{AC}^2+\frac{\tau_{ABC}^2}{4}]^{\frac{1}{2}},
\end{eqnarray}
where, both inequalities are based on the arithmetic geometric mean
inequalities $a^2+b^2 \geq 2ab$ for $a, b \geq 0$. The second
equality is due to (\ref{zh1}), the third and last equalities are from the results in
\cite{vcjk}, and the tangle $\tau_{ABC}=4\lambda_1\lambda_2=4\mu_1\mu_2$.

Inequality (\ref{zh2}) is tighter than the inequality
$C_{A|BC}^2\geq 2C_{AB}C_{AC}$ obtained by arithmetic geometric mean
inequality from CKW type. The result consists with the one from the
summation form only when $\tau_{ABC} = 0$. Therefore,
inequality(\ref{zh2}) is basically different from the summation-form
monogamy inequalities.

For arbitrary quantum correlations $Q$, the multiplication form of
monogamy relations may be not as tight as (\ref{zh2}). However, when
$\alpha=2$ in (\ref{eq1}), we can directly get
\begin{eqnarray}\label{eq3}
\sqrt{Q(\rho_{AB})Q(\rho_{AC})}\leq
\frac{\sqrt{2}}{2}Q(\rho_{A|BC})\leq Q(\rho_{A|BC}).
\end{eqnarray}
Namely, any monogamy inequalities (\ref{eq1}) satisfied by a quantum correlation $Q$ imply
a product form (\ref{eq3}).
This is also true for multipartite case.
From a monogamy relation
$$
Q^2(\rho_{A|B_1B_2\cdots B_{n-1}})\geq \sum_{i=1}^{n-1}Q^2(\rho_{AB_i}),
$$
we obtain
$$
\sqrt[n-1]{\prod Q^2(\rho_{AB_i})}\leq \frac{\sum_{i=1}^{n-1}Q^2(\rho_{AB_i})}{n-1} \leq \frac{Q^2(\rho_{A|B_1B_2\cdots B_{n-1}})}{n-1},
$$
i.e.
$$
\sqrt[n-1]{\prod
Q(\rho_{AB_i})}\leq\frac{Q(\rho_{A|B_1B_2\cdots
B_{n-1}})}{\sqrt{n-1}}.
$$

In \cite{pkac,ppam,scam1,scam2}, the authors studied some
restrictive relationships for different quantum correlations such as
contextually \cite{pkac}. For instance, for general three-qubit
states, the Bell nonlocality and three tangle have the
complementarity relation
$\max\{M_{\rho_{AB}},M_{\rho_{AC}},M_{\rho_{BC}}\}+\tau \leq 2$
\cite{ppam}. The internal entanglement of a bipartite system, and
its correlations with an environment system, have the relation
$E(A_1:A_2)+E(A_1A_2:B)\leq E_{\max}$ \cite{scam1}. All these
inequalities in summation form can be transformed into product form
in a similar way.

\section{conclusion and discussion}

We have studied the monogamy relations in product form for Bell
nonlocality and concurrence of multi-qubit states. Product-form
monogamy relations can be obtained from the summation form by
arithmetic geometric mean inequality, like inequalities (7) and
(12). Such product-form monogamy relations can be tighten by
tightening the summation-form monogamy inequalities or by using
weighted arithmetic geometric mean inequalities. Nevertheless, the
inequalities (8), (10) and (15) are not obtained from the arithmetic
geometric mean inequalities. They are tighter than the ones derived
from the summation-form relations. These product-monogamy relations
may be of their own applications in quantum information processing.
For instance, the inequality (8) can be used in three point quantum
network. Thus, for multipartite quantum nonlocality the product-form
monogamy relation may be applied to more complex quantum networks
\cite{atmo,mcrr}. Our results may highlight further researches on
other quantum correlations like quantum discord and quantum
steering.

\bigskip
Acknowledgments: This work is supported by the NSF of China under
Grant Nos. 11861031 and 11675113,
the Key Project of Beijing Municipal Commission of Education (Grant No.
KZ201810028042), and Beijing Natural Science Foundation (Z190005).

\end{document}